# Issues with the High Definition Space Telescope (HDST) ExoEarth Biosignature Case

A Critique of the 2015 AURA Report
*"From Cosmic Birth to Living Earths: the future of UVOIR Astronomy"*

Martin Elvis
*Harvard-Smithsonian Center for Astrophysics*
25 September 2015

## Executive Summary

The AURA report "*From Cosmic Birth to Living Earths*" advocates a 12-meter diameter optical/near-IR space telescope for launch ~2035. The goal that sets this large size, roughly double that of the James Webb Space Telescope (JWST), is the detection of signs of life (biosignatures) from Earth-like planets in their habitable zones around G-stars. The discovery of a single instance of life elsewhere in the universe via an exobiosphere detection would be a profound event for humanity. But not at any cost. At a cost of $8B - $9B this *High Definition Space Telescope* (HDST) would take all the NASA astrophysics budget for nearly 20 years, unless new funds are found. For a generation NASA could build no "Greater Observatories" matching JWST across the rest of the spectrum. This great opportunity cost prompted me to investigate closely the driving science case for HDST – exobiosphere detection.

My conclusions are that: (1) The focus on G-stars is not well justified; (2) only G-stars require the use of direct imaging; (3) in the chosen 0.5 – 2.5 µm band, the available biosignatures are ambiguous and a larger sample does not help; (4) the expected number of biospheres is 1, with a 5% chance of zero; (5) the accessible sample size is too small to give a 3 sigma upper limit that would show that exobiospheres are rare; (6) to get a sufficiently large sample would require a much larger telescope; (7) the extraordinarily rapid progress in the spectroscopy of planets around M-stars - both now and with new techniques, instruments and telescopes already planned - means that a biosignature will likely be found in one before HDST could complete its search in ~2045.

For all these reasons I regretfully conclude that HDST, while commendably ambitious, is not the right choice for NASA Astrophysics at this time.

The first exobiosphere discovery is likely to be such a major event that scientific and public pressure will produce new funding across a range of disciplines, not just astrophysics, to be brought to bear on the nature of Life in the Universe. Then will be the time when, armed with much more knowledge than we now have, a broader science community can advocate for a mission that will make definitive exobiosphere measurements.

## 1. The proposed High Definition Space Telescope (HDST)

The August 2015 AURA report "*From Cosmic Birth to Living Earths*"[1] (CB2LE) advocates a 12-meter diameter space telescope for the next large NASA astrophysics mission, the "High Definition Space Telescope" (HDST). There is no doubt that the HDST goal of discovering biosignature gases in the atmospheres of exoEarths in a search for life beyond the Earth is a self-evidently worthy goal for astrophysics, and for humanity in general. But not at any cost. In the NASA context, the opportunity cost to the rest of astrophysics is likely to be huge.

HDST would be a flagship mission, and would launch around 2035[2] or later[3]. HDST would have a primary mirror roughly double the diameter of the $8.7 billion[4] *James Webb Space Telescope*[5] (JWST). HDST would cost a similarly large amount. At the CB2LE rollout event at the American Museum of Natural History, Mark Postman is quoted as giving a cost " *'on the order of the James Webb telescope', which he later specified as being between $8 billion and $9 billion.*"[6]. The total NASA astrophysics budget for the next 10 years is $5 billion, according to Paul Hertz, the NASA Astrophysics Division Director[7]. A mission in the $10B class would thus take all of the available funding for nearly 20 years, ruling out a balanced program. Barring a major increase in the NASA astrophysics budget or major non-US participation, NASA astrophysics would become a single telescope operation for a generation.

It is possible that a single telescope could be so important that it justifies monopolizing NASA astrophysics for a generation. But the case must then be extremely strong. We must consider what a series of more modest, yet still extremely powerful, missions would be foregone by this choice and what they could have accomplished. HDST would span one decade of the spectrum (0.1 – 2.5 μm, Table 5.3). The NASA "Great Observatories" now operating span total of roughly 4 decades of the electromagnetic spectrum at comparable sensitivity: *Chandra* in the X-rays, *Hubble* in the UV to near-IR and *Spitzer* in the longer wavelength. (The original cryogenic *Spitzer* plus the ESA far-IR telescope *Herschel*, extended this well-matched coverage another 2 decades.) This combination has enabled revealing pan-chromatic studies of all classes of astrophysical objects and rapid follow-up of discoveries in one band with observations in the others. This synergy between the many new windows opened onto the cosmos, it could be claimed, is a major reason for our current Golden Age of Astronomy. NASA has initiated studies of a series of "Surveyor" missions[8] that would constitute the next generation "Greater Observatories", if they can be launched. HDST would appear to require giving up on sensitive broad spectral coverage[9]. Closing these windows is a strategic decision for the field of the first order.

---

[1] URL: http://www.hdstvision.org/report
[2] http://static1.squarespace.com/static/558adc44e4b002a448a04c1a/t/55ce4b1fe4b0bd9475500032/1439583007089/SPIE_Delivered_Aug2015_compressed.pdf
[3] If the program is limited to the current $5B/decade NASA astrophysics budget then, for a WFIRST-AFTA launch in 2025, HDST launch is likely to be around 2045.
[4] http://www.space.com/12759-james-webb-space-telescope-nasa-cost-increase.html
[5] http://www.jwst.nasa.gov
[6] Calla Cofield, Space.com, http://www.space.com/29878-alien-life-search-hdst-space-telescope.html. New York Times science reporter Dennis Overbye puts HDST in the *"$10 billion class"* http://www.nytimes.com/2015/07/14/science/space/the-telescope-of-the-2030s.html?_r=0
[7] http://files.aas.org/head2015_workshop/HEAD_2015_Paul_Hertz.pdf, integration of slide.14

The case for HDST then has to be extraordinary. The result will need to be definitive - either a strong detection of life elsewhere, or a series of significant non-detections that mean life is rare even on exoEarths in habitable zones. This case is examined in this paper.

It is not the custom in our field to directly critique the mission proposals of others, although we critique each other's science papers closely. But the scale of HDST and hence the opportunity cost it would impose seems to require due diligence by the community.

The large mirror surely drives the cost of the mission. Let us then look closely at the assumptions that lead the report to a 12-meter diameter primary mirror for HDST. Half of the science justification (chapter 4) is based on astrophysics, from the formation of the first galaxies to star formation in our Galaxy. The other half (chapter 3) is devoted to finding habitable Earth-like planets and searching for biosignatures in their atmospheres. It is the requirements for angular resolution and sensitivity to biosignatures that drives the 12-meter requirement (p.89, 95).

All the CB2LE arguments from Chapter 3 for a 12-meter telescope based on exoEarth biosphere searches are discussed below. Other factors - the rest of astrophysics, the technical readiness of the design and the cost control concepts, though extensively discussed in CB2LE, are not addressed here.

## 2. Why G-stars?

The decision of CB2LE to focus HDST solely on searches of G-type stars for exoEarths is fundamental to the entire HDST biosignature case. Yet the motivation for this choice is left unstated. CB2LE simply asserts "*The long-term future is to go beyond M-dwarf stars to focus on Solar-type stars*" (p.25). Why are Solar analog G-dwarf stars the long-term future? Is finding an Earth around a yellow sun more like us and so more interesting? If so, is that enough? If not, what is the justification?

---

[8] One of which is a "Habitable-Exoplanet Imaging Mission", http://spacenews.com/nasa-seeks-astronomy-missions-concepts-for-2020-report/

[9] Other agencies have no matching missions planed that fill the gaps: SPICA might have done so in the far-IR but was dropped in late 2013 (http://sci.esa.int/cosmic-vision/53635-spica/); no comparable UV mission is under development, though HDST is intended to fill this gap - an analysis of how well it does so would be worthwhile; ESA's Athena X-ray mission, while having large area, will only have 5 arcsec resolution, 10× worse than Chandra, and so will be source and identification confused and background limited well before the deep survey levels of the *Chandra* and *Hubble* Deep Fields.

Pinning the mission on G-stars has enormous cost. CB2LE state that the highly productive transmission transit spectroscopy technique now being applied to G-M stars will not work on Earth analogs orbiting Solar analog stars. As a result HDST needs to use direct imaging of these planets. To directly image small planets near their stars requires a large aperture to produce a small diffraction limit of ~0.05 arcseconds, so that a planet at 1 AU can be separated from its star out to ~15 parsecs. It also requires coronography in some form to suppress the glare of the system's starlight by 10 orders of magnitude, which is clearly technically challenging.

The dominant scientific priority of G-stars is not obvious. Recent analyses using the Kepler mission planet census reveal that there are twice as many small planets around M-stars as G-stars (Mulders et al. 2015). The spectral differences between M and G-stars are not so large (T ~ 4000 K vs 6000 K). The formation history of planetary systems might depend strongly on the central star mass, but only a factor ~2-3 in mass separates M-stars from the Sun. Unless there is a strong theoretical expectation that the stellar mass or spectrum will have a large effect on the potential for life, why should we demand extraordinary performance from our telescopes to look at G-stars given the great cost of doing so, and the relative ease of using M-stars? (see Section 3.)

It is quite possible, even likely, that a biosignature will be found from an exoEarth around an M-star well before HDST could be built (section 3). Would this not be a great discovery? Any single biosignature tells us that life is not unique to Earth, a profound event in human history. By comparison, whether the signature is found from a plane around a G-star or an M-star is a minor, albeit interesting, difference.

## 3. Why Imaging?

The cost and complexity of directly imaging exoEarths around G-stars mean that other techniques must be shown to be definitively inadequate.

The current primary technique for exoplanet detection is to use exoplanet transits in front of the host star. Transits also yield planet sizes that are needed to measure their densities while direct imaging does not. The transit technique was used to great effect by the NASA *Kepler* mission, which has found over a thousand confirmed exoplanets, nearly 4000 candidate planets, including nearly 1000 Earth-sized candidates, 14 of them in the habitable zone (July 2015[10]). Kepler stars are too faint though for atmospheric feature detection.

However, the NASA TESS mission[11] will survey ~200,000 bright, nearby stars for exoplanets beginning in 2017/2018. Roughly 5 exoEarths are expected from the 2-year nominal mission (Sullivan et al., 2015). The ESA PLATO mission will make a more extensive survey of ~1,000,000 stars from 2024-2030[12]. Of order two dozen transiting exoEarths will then likely be known well before HDST flies.

---

[10] http://exoplanetarchive.ipac.caltech.edu/docs/counts_detail.html
[11] http://tess.gsfc.nasa.gov

The smaller M-stars give a stronger transit signal, and so searches for biosignature gases for super-Earths around M-stars are already underway from the ground. Major advances can be expected using the bright TESS and PLATO exoEarths and major new instruments including JWST and the 30-meter class telescopes. For example, the Doppler-shift technique of Snellen et al. (2010) appears to be suitable for searching for atmospheric signatures of exoEarths with the upcoming generation of Extremely Large Telescopes (E-ELT, GMT, TMT), as explored by Snellen et al. (2013), and Rodler & Lopez-Morales (2014).

It is quite plausible then that biosignatures will be found in an exoEarth in the habitable zone of a M-star before HDST is launched. Were HDST to add a single biosignature exoEarth around a G-star about 10 years after launch (i.e. ~2045, see section 6 below), while interesting, would not be game-changing.

CB2LE notes that the transit technique only works for the subset of planetary systems that are roughly edge-on to us, while imaging can find these missing systems. Why does it matter, though, if we miss the more face-on systems? They are not going to be physically different. So long as our sample is large enough we can do the statistics just as well with the edge-on systems and correct them for the geometry. Transits are then perfectly adequate. Moreover the Snellen et al. (2010) technique is already being extended to non-transiting planets (Brogi et al., 2012). HDST itself would have the inverse issue as it would be less efficient at detecting more edge-on exoEarths which appear too close to their stars for more of their orbits as their inclination is increased.

**4. Ambiguous Biosignatures**

Several biomarker signatures in the optical/near-IR band (0.5 - 2.4 μm) (Figures 3-8, 3-9) are discussed by CB2LE. The problem is that abiotic origins for individual biomarkers, e.g. oxygen and ozone, are being discovered (Domagal-Goldman et al. 2014, Wordsworth and Raymond 2014), so that a single molecule does not provide a robust signature. Acknowledging this issue, CB2LE concludes "*Even with superb data, there is no single "smoking gun" biosignature gas… This means that aiming for a robust detection of biosignature gases on a single planet may not be enough… a large number of candidate exoEarths with detected biosignature gases will be needed*" (p.27). But if any individual biosignature gas is insufficient, how does a large sample, each with ambiguous biosignatures, help? Several different biomarker gases must be found in *each* candidate in order to rule out non-biological causes. What combination would be convincing? Is this measurable? CB2LE does not say. No doubt much research has yet to be done to determine the answer.

---

[12] http://sci.esa.int/plato/

Earth's atmosphere over the 0.5 - 2.4 μm band discussed in Chapter 3 contains half a dozen different gases (Figure 3-8). This appears promising. But later (p.37) CB2LE notes that HDST performance beyond 1 micron will be severely degraded for exoEarth spectroscopy (section 5). The remaining optical (0.5 - 1 μm) band contains only 2 different gases in any single Earth-like planet (Figure 3-9). It seems unlikely that this enough to produce a strong result. If the near-IR is essential to get sufficient biomarkers, as seems likely, this has a big effect on the sample size and on mirror size, and so on the case for HDST (section 5).

Why does CB2LE limit itself to the optical/near-IR (0.5 - 2.5 μm) band for biosignature searches? Other signatures are present out to at least 16 microns, and shortward into the UV. Bétrémieux and Kaltenegger (2015) show that the UV contains valuable diagnostics. Sagan et al. (1993) show a spectrum of Earth from the *Galileo* spacecraft with strong features from 4 – 5 μm, including $CH_4$, $N_2O$, CO. True, the inner working angle (IWA) for a 12-meter would not allow direct imaging at these wavelengths. Other approaches though are being studied. The EChO transmission spectroscopy mission targeted features out to 11 μm and would have gone to the 16 μm $CO_2$ feature had the detector technology been ready (Tinetti et al. 2015). ARIEL, a mission derived from EChO, is a candidate for the ESA M4 Mission, which is planned to launch in 2025[13].

## 5. ExoEarth Spectra

CB2LE says that "*Spectroscopy spanning from visible wavelengths to 2.4 microns is desired so that multiple spectral bands of each molecular species can be observed…*" (p.36/37). As we saw in section 3 above, the optical band alone has features from only 2 gases in its spectra for any given planet (Figure 3-9). Given the ambiguity of optical spectra alone it appears that near-IR spectra are required for a clear exo-biosphere detection.

However, CB2LE also states that "*Reaching to near-IR wavelengths presents challenges*" (p.37). This is due to the diffraction limit of the telescope, and so the IWA, growing linearly with wavelength. "*The yield of planets with near-IR spectroscopy will therefore be lower than for optical spectroscopy*" (p.37). Figure 3-16 quantifies this. For 12-meter mirror the optical band gives 50 - 80 exoEarth candidates (for IWA= 3.6 and 2 λ/D respectively), while the near-IR gives ~30[14]. Given this issue CB2LE backs off on the contrast ratio requirement in the near-IR by a factor 10 (from $10^{-10}$ to $10^{-9}$, p.37). Figure 3-6 indicates that this change puts true Earth twins out of reach of HDST. This requirements change is not discussed further. The more consistent conclusion is that HDST will find 20 exoEarths that can be searched for multiple biosignatures.

## 6. ExoEarth Sample Size and Yield

---

[13] ARIEL: The Atmospheric Remote-Sensing Infrared Exoplanet Large-survey, ariel-spacemission.eu
[14] Figure 3-16 assumes $\eta_{Earth}$ = 0.2 which may be optimistic – see Section 6.

"*The HDST report recommendation is to seek dozens of exoEarths for detailed atmospheric characterization*" (p.29). Here "exoEarth" means an Earth-sized planet in the habitable zone of its G-star. Expanding on this, Section 3.8 Summary (p.43) begins by stating: "*HDST's primary goal is to detect and spectroscopically examine dozens of exoEarths in their star's habitable zone. To find these gems, the telescope will survey many hundreds of nearby main-sequence stars.*" The report adopts a goal of 95% confidence of finding at least 1 exoEarth with biosignatures.

How many exoEarths would HDST find? The effect of changing the assumed $\eta_{Earth}$ (0.05 – 0.2), the exozodiacal ("exozodi") background (3 - 100 zodi, the Solar System value), the IWA (2 – 3.6 $\lambda/D$) and the mirror diameter (2 – 20 m) are investigated in Figure 3-15. For the minimum exozodi (3 zodi) the yield of exoEarth candidates provided by a 12-meter primary ranges from ~12 to ~75 in 1 year of on-sky observations. If the exobiosphere program is ½ of a 7-year program, allowing 3 years for spectroscopic follow-up (see Section 7) then this translates into ~40 - ~260 exoEarths. The largest assumed exozodi would cut these values by a little more than 50%; I will assume the minimum considered 3 zodi level.

To design a robust program, which values should be assumed? The pessimistic IWA is probably required, as near-IR spectra benefit from this (p.37, Section 5); $\eta_{Earth}$ = 0.1 works for M-stars (Dressing & Charbonneau 2015) and small planets seem to be a factor 2 rarer for G-stars (Mulders et al. 2015), so taking the pessimistic $\eta_{Earth}$ = 0.05 seems appropriate. A total number of exoEarths closer to 40 than to 260 thus seems more likely for planning purposes at this point.

Of the detected exoEarths some fraction, $\eta_X$, would have detectable biosignatures. If $\eta_X$ = 0.1 then Figure 3-10 shows that 30 exoEarths are needed to yield at least one biosignatures (at 95% confidence, 2 sigma equivalent). Finding the first exo-biosphere will surely be an epochal event. We will then know that there is life elsewhere. (But see Section 4.) An expectation of a single detection though is a shaky basis for a flagship mission. Zero is also quite possible (at the 5% level). As $\eta_X$ is just an assumed value zero may even be more likely.

If no exobiospheres were found, the resulting 95% limit of $\eta_X$ <0.1 is not a strong result. It does not say whether exoEarths with biosignatures are rare or common. If we demand the more commonly used, and more definitive, 3 sigma upper limit then, in the case that no biosignatures are found, the same 30 spectra would put a limit of $\eta_X$ < 0.2 (Fig.3-10), i.e. no more than than 1 in 5 exoEarths have biosignatures. With this limit exobiospheres could well be common without HDST finding one.

A more robust approach is to scope HDST so that a *non-detection* is itself a noteworthy event – a clear statement that biospheres are rare and that the Earth is special. Instead the criteria adopted in the report are quite weak. Pushing $\eta_X$ to < 0.01 would be interestingly restrictive. Other biospheres would then be quite rare. But Figure 3-10 shows that this goal would require 300 exoEarth spectra for a 2 sigma (95%) result, or 600 spectra for a 3 sigma (99.7%) result. If 30 is a plausible HDST program (Section 7), then 600 would require a 20× faster rate. If the supply of stars could remain as bright as the first 30 then this would translate into a 53-meter diameter primary mirror, but this is not possible[15]. (As the stars would be ~3× further away, a 36-meter mirror is in any case needed to maintain the same resolution in AU at the planetary system.) Such structures are more than we can reasonably hope to build in the next 20 years, without a revolution in spacecraft manufacturing.

It is interesting to compare the criteria adopted for HDST with the measured value of the exoEarth occurrence rate, $\eta_{Earth}$, around M-stars. CB2LE states that this value is only known with "*large uncertainties*" (p.32) as it is constrained to lie between 0.09 and 0.31 (Dressing & Charbonneau 2015). Had this result been an upper limit of order 0.1 then we would indeed not have learned much, as pointed out above for the HDST program. However, a positive measurement is a huge advance that means that exoEarths are undoubtedly common, at least around M stars. Even 10× smaller error bars would now be only a (valuable) refinement, not a fundamental result.

To get this result Dressing & Charbonneau (2015) examined 156 planet candidates from *Kepler*. CB2LE argues that 30 stars would be sufficient to determine the G-star $\eta_{Earth}(G)$ (Section 3.4). These do not appear to be consistent criteria. If we use 150 as a minimum sample, then to determine $\eta_{Earth}$ as well as for M-stars in the same observing time requires a 26-meter mirror.

## 7. Exposure Times, Program Size

The total exposure time that would be needed for the HDST exo-biosphere program is worth estimating.

For the imaging program to discover exoEarths Figure 3-17 provides the information that 3-band imaging of a Solar System twin at 13.5 parsecs would take 40 hours for 3 bands, each of 10% spectral width. Hence[16] "many hundreds" of stars would take ~600 × 40 hours, which is 1000 days, or 3 years at 90% efficiency (a high but plausible value). ExoEarth candidates will then presumably have to be re-observed at least once some months apart to confirm that they are planets, not background sources; one follow-up band adds a year of observing time. As exoEarth biosignatures are half the science case for HDST such a large program is reasonable and would be completed ~8 years after launch, if observing time is shared equally with the rest of astrophysics.

---

[15] In practice to get 600 G-type stars requires a maximum distance 2.7× larger than to get 30, so they would be ~2 magnitudes (a factor 7.3) fainter. The resulting primary mirror diameter requirement would then become 143 meters.)

[16] Assuming 13.5 pc is the mean exposure time distance. Fig 3-14 shows that a radius of ~35 pc is needed to reach 582 stars. Half the volume is reached at 28 pc. At 28 pc a star is 4.3 × fainter and so would need ~172 hr of imaging rather than 40hr. A 4 × longer program would take 12 years of 100% of HDST observing time. (Compared with Fig. 3-17 the planets would be 2.6× closer to the star, which would be at the limits of detectability.)

If the sample size critique above (section 4) is correct though, a sample 10-20 times larger would be needed. This rapidly becomes implausible without a significantly larger primary mirror to shorten exposure times[17].

No exposure times are given for exoEarth spectroscopy in the report. As spectra divide the light received into smaller wavelength pieces, spectroscopy typically takes longer than imaging. CB2LE tells us "*For internal coronagraphs… a sequence of observations each covering a 10-20% based pass would cover the full desired range*" (p.37). So ~10 exposures are needed, rather than 3 for imaging. If we require 1% spectral resolution (minimal, but adequate, given the width of the features in Figure 3-8) then each exposure needs to be 10× longer than for imaging. So each full spectrum would take 33 times longer than the imaging for that star system. I.e. about 1300 hours or 38 days[18]. For a sample of 20 spectra this totals 2.3 years at 90% efficiency.

Including the precursor imaging program (see section 6) 6.3 years of observing time would be needed to carry out the complete HDST exoEarth biosignature program. For a 10-year mission typical of NASA flagships this is a substantial time. It is roughly in keeping though with the 50% of the HDST science case that is exoEarth biosignatures.

Any increase in sample size or spectral resolution requirements would rapidly lead to mission-length, >10 year, observing programs.

## 8. Conclusions

"*Extraordinary claims require extraordinary evidence.*" Carl Sagan's words apply just as strongly to proposed flagship missions as to scientific results. The case for such a costly mission as HDST must be extraordinary. It is not obvious that this is so for the HDST exoEarth biosignatures case as presented in CB2LE.

The arguments presented here suggest that whether HDST will be the telescope that produces the first signature of life beyond the Earth is doubtful. *First*, it seems likely that the biosignatures accessible to HDST will be ambiguous; *Second*, even if they are definitive, the HDST G-star exoEarth program has a ~5% chance of finding zero exo-biospheres, even if 1 in 10 exoEarths have biospheres, and would not set a strong limit on how rare biospheres are around G-stars. *Third*, this result will come around 2045, 10 years after a plausible 2035 launch[19]. By then HDST could well have been scooped by the discovery of biospheres beyond the Earth from ground-based telescopes, including the ELTs.

---

[17] Bringing mission costs down by large factors is desirable, not impossible, and may well be timely. It should be the subject of another vigorous discussion within and between all the communities using space for science.

[18] An Earth will move 38 degrees around its orbit in this time. This does not seem to be discussed.

[19] See footnote 2, and assuming 50% of the time goes to the non-exoEarth astrophysics program.

Strengthening the CB2LE plan to show that, in the case of no detection, exobiospheres are rare (1% limit at 3 sigma confidence) leads to 50-meter class telescopes being needed. At such large mirror diameters even an enthusiast for deriving requirements entirely from the demands of science would likely admit that issues of technology, cost and schedule had to be considered.

Of the 13 references used for this paper, 9 are from 2014 or 2015, showing how rapidly the exoEarth field is developing. It is probable that the assumptions made for HDST will change within a few years. This rapid pace alone argues that any space mission for exobiosphere and/or exoEarth detection must be robust against change. HDST does not appear to be so.

Given the minimal robustness of the exobiosphere arguments for a 12-meter HDST, and the opportunity cost of limiting NASA Astrophysics to this one flagship mission for 20 years, I regretfully conclude that HDST - though commendably ambitious - is not the right choice for NASA astrophysics at this time. If the program could be carried out for ~20% of the likely cost, as intended for the "*Habitable-Exoplanet Imaging Mission*"[8], then the opportunity cost would be much reduced and the calculus would be quite different.

Most astronomers would probably agree with CB2LE when it says *"…keep[ing] flagship missions as an active part of NASA's balanced program maintains the focus on projects of sufficient vision and scope that they 'make the pie bigger'."* A balanced program needs more than one flagship mission. To do HDST and the "Greater Observatory" flagships needs more than a doubling of the NASA astrophysics budget, or another source of funds. How could this happen?

The discovery of another biosphere from observations of M-star exoEarths would surely create a sensation. Demand, both scientific and public, for energetic and intense follow-up would be strong. That will be the time to push for additional new funding for programs to learn more about life in the universe, without the burden falling solely on the rest of astrophysics. The study of Life in the Universe goes beyond astrophysics and that funding could be sourced from elsewhere including biology, geology, chemistry, and planetary science. Or, given the felt importance, from entirely new funds. In this larger picture $2B/year to build a telescope to search for exobiospheres in a timely fashion would not be a big request.

The future of searches for life in the universe in the broader astrophysics context is clearly controversial. The community should debate thoughtfully whether HDST passes the Sagan test.